\documentstyle[12pt]{article}
\title {\bf{Quantum dynamics of N=1, D=4 supergravity chiral compensator}}
\author{I.L.Buchbinder\thanks{e-mail: josephb@tspi.tomsk.su}\\
      {\small\it{Department of Theoretical Physics}}\\
      {\small\it{Tomsk State Pedagogical University}}\\
      {\small\it{Tomsk 634041, Russia}}\vspace{2mm}\\
               A.Yu.Petrov\\
       {\small\it{Department of Quantum Field Theory}}\\
       {\small\it{Tomsk State University}}\\
        {\small\it{Tomsk 634050, Russia}}}
\date{}
\begin{document}
\maketitle
\thispagestyle{empty}
      Short title: Quantum dynamics of chiral compensator

       PACS numbers: 04.60, 04.62, 11.30
\newpage
\thispagestyle{empty}
\begin{abstract}
A new four-dimensional $N=1$ superfield model is suggested. The model
is induced by supertrace anomaly of matter superfields in curved
superspace and leads to effective theory of supergravity chiral compensator.
A renormalization structure of this model is studied, one-loop counterterms
are calculated and renormalization group equations are investigated. It is
shown that the theory under consideration is infrared free.
\end{abstract}
\newpage
\pagenumbering{arabic}
{\section {Introduction}}

  It is known that $N=1$, $D=4$ supergravity can be formulated as a dynamical
theory of curved $N=1$ superspace in terms of vector superfield $H^m(z)$ and
compensators: chiral $\Phi(z)$ and antichiral $\bar{\Phi}(z)$ [1-3] (see also
 [4-7]). Here $z^M$ = $(x^m, \theta^{\mu},\bar{\theta}^{\dot{\mu}})$ are the
superspace coordinates. The different aspects of superfield supergravity were
discussed by many authors but quantum aspects have not been investigated so far
in details. The general approach suggested in refs. [8,9] was not further
developed and applied to concrete problems apparently because of rather
complicated structure of superfield supergravity itself and
some unclear questions of quantization.

   In this paper we consider a simplified model of $N=1$, $D=4$ superfield
supergravity, where the only dynamical fields are the compensators
$\Phi(z)$ and $\bar{\Phi}(z)$ in flat superspace. We show that this model
allows
to carry out a detailed investigation of quantum aspects, possesses remarkable
properties in infrared domain and provides a possibility to study the questions
of infrared behaviour in superfield supergravity.

   The model under discussion is based on an idea of induced gravity. This idea
was developed in context of string theory [10], was investigated by many
authors
(see f.e. [11-17]) and has led to theory of two-dimensional dilaton gravity.

   Four-dimensional model of quantum gravity induced by conformal anomaly
of matter fields has been developed in [18]. This theory is based on action
generating conformal anomaly [19,20] and describing dynamics of conformal
factor
of metric. It was proved [18] that such a model is super-renormalizable in
infrared limit, possesses infrared stable fixed point and provides a physically
acceptable mechanism for computation of cosmological constant. A further
development of the model was carried out in [21-33], where a number of
questions
connected with influence of torsion field, phase transitions on curvature,
possible modification of Newton law of gravity, higher dimensions
generalization, construction of general four-dimensional dilaton gravity model
have been investigated.

  In order to formulate an analogous approach in superfield supergravity we
should answer at least two questions: what is a structure of supertrace anomaly
in N=1 curved superspace and what is the form of superfield action generating
this anomaly? Both answers are known at present. General answer for the first
question is given by refs. [34,35], answer for the second one is given by ref.
[36]. If we add the action of supergravity with cosmologocal term to the
action,
generating anomaly, and put vector superfield $H^m$ to be equal to its value in
flat superspace, we will obtain an effective model of superfield supergravity
where dynamical variables can be only the compensators $\Phi$, $\bar{\Phi}$. As
a result we get a new model of interacting chiral superfields in flat
superspace. We want to stress specially that this model is completely
formulated
in superfield terms although it is essentially different from standard Wess-
Zumino model due to presence of higher derivatives.
\footnote{We do not discuss here the question of unitarity of $S$-matrix,
caused
by higher derivatives. The model under consideration should be understood as
an effective theory. Both model introduced in ref. [18] and our model are only
aimed for investigation of low-energy behaviour.}.

  The paper is organized as follows. In section 2, a formulation of the model
is
given, a structure of supertrace anomaly generating superfield action is
discussed and the propagators an vertices are written down. Section 3 is
devoted
to classification of divergences and possible counterterms. In section 4, an
explicit calculation of one-loop counterterms is carried out. Section 5 is
devoted to renormalization group analysis of effective couplings. In Appendix a
component form of the lagrangian of model under consideration is given.

\vspace{5mm}
{\section{Formulation of model}}
   We start with brief review of superconformal anomaly and action generating
this anomaly. The basic objects in terms of which superconformal anomalies are
formulated are the superfields $V_{\alpha\dot{\alpha}}$ and $V$, called a
supercurrent and supertrace respectively [37,38] (see also [7]) and satisfying
the conservation law
%%%%%%%%%%%%%%% (1) %%%%%%%%%%%%%%%%%%%
\begin{eqnarray}
\bar{D}^{\dot{\alpha}}V_{\alpha\dot{\alpha}}=\frac{1}{6} D_{\alpha} V\\
\bar{D}_{\dot{\alpha}} V=0\nonumber
\end{eqnarray}

   Let us consider a class of superconformally invariant theories [39,40,7].
In this case the classical supertrace $V$ is equal to zero. However in quantum
theory the superconformal anomaly can arise and as a result we get the anomaly
supertrace $V_A$ in the form
\begin{equation}
\label{trace}
V_A=-{(4\pi)}^{-2}(aW^2 +bG+c(\bar{D}^2 -8R)D^2R)
\end{equation}
where $a$, $b$, $c$ are the numbers depending on the field content of the
model,
$W^2=W^{\alpha\beta\gamma}W_{\alpha\beta\gamma}$, $R$, $G_a$,
$W_{\alpha\beta\gamma}$ are superfield strengths expressed in terms of
supergravity prepotentials $H^m,\Phi,\bar{\Phi}$;
$G=W^2-\frac{1}{4}(\bar{D}^2-8R)(G^a G_a+2R\bar{R})$- is the superspace density
of the Gauss- Bonnet superinvariant,
$D_A=(D_a, D_{\alpha}, \bar{D}_{\dot{\alpha}})$
are supercovariant derivatives (we use the denotions accepted in ref. [7]).

   The action generating anomaly (\ref{trace}) has been found in ref.[36].
Being transformed to conformally flat superspace this action takes the
following
form

\begin{eqnarray}
\label{act}
\Gamma_A&=&\frac{1}{16\pi^2}\int d^8 z
(8c\partial_a\bar{\sigma}\partial^a\sigma +\\
&+&\frac{1}{32}(32c-b)\bar{D}^{\dot{\alpha}}\bar{\sigma}D^{\alpha}\sigma
(\partial_{\alpha\dot{\alpha}}(\sigma+\bar{\sigma})-
\frac{1}{2}\bar{D}_{\dot{\alpha}}\bar{\sigma} D_{\alpha}\sigma))\nonumber
\end{eqnarray}
Here we have used the "flat" supercovariant derivatives
$D_{\alpha}$, $\bar{D}_{\dot{\alpha}}$, $\partial_{\alpha\dot{\alpha}}$;
and $\sigma=\ln\Phi$, $\bar{\sigma}=\ln\bar{\Phi}$.

Let us also consider the action of N=1 superfield supergravity, in conformally
flat superspace we obtain

\begin {equation}
 S_{SG}=-\frac{m^2}{2}\int d^8 z \Phi\bar{\Phi}
+[\Lambda\int d^6 z \Phi^3 +h.c.]
\end {equation}
where $m^2=\frac{6}{\kappa^2}$, $\Lambda$ is the cosmological constant and
$\kappa$ is the gravitational constant. We will investigate the theory action
of which is a sum of the actions (3) and (4)
   Denoting
\begin{eqnarray}
\frac{Q^2}{2}&=&8c\nonumber\\
\xi_1&=&\frac{1}{32{(4\pi)}^2}(32c-b)\\
\xi_2&=&-\frac{1}{64{(4\pi)}^2}(32c-b)\nonumber
\end{eqnarray}
we get a superfield theory in flat superspace with the action of the form
\begin {eqnarray}
\label{action}
S&=&\int d^8z(-\frac{Q^2}{2{(4\pi)}^2}\bar{\sigma}\Box\sigma+
\bar{D}^{\dot\alpha}\bar{\sigma}
D^{\alpha}\sigma\times\nonumber\\
&\times&({\xi_1\partial_{\alpha}}_{\dot{\alpha}}(\sigma+\bar{\sigma})
+\xi_2\bar{D}_{\dot{\alpha}}\bar{\sigma} D_{\alpha}\sigma)-
\frac{m^2}{2} e^{\sigma+\bar{\sigma}})+(\Lambda\int d^6z e^{3\sigma}+h.c.)
\end {eqnarray}
The $Q^2, \xi_1, \xi_2, m^2, \Lambda$ will be considered as the arbitrary and
independent parameters of the model. We will call the model with action
(\ref{action}) the theory of supergravity chiral compensator.
Namely this theory is a basic object of investigation in this paper.

   In order to calculate the counterterms and to find the divergences we should
study a structure of supergraphs of the theory.The strucure of supergraphs
is defined by a form of propagators and vertices. In its turn propagators and
vertices can be found by standard way on the base of action (\ref{action})
(see this procedure for example in ref.[7]).

The theory under consideration is characterized by a matrix propagator
\begin {equation}
G(z,z')=\left(
\begin{array}{cc}
   G_{++}(z,z') & G_{+-}(z,z')\\
   G_{-+}(z,z') & G_{--}(z,z')
\end{array}
\right)
\end {equation}
  satisfying the equation
\begin {equation}
\label{eq}
\left(
\begin{array}{cc}
  9\Lambda & (\frac{Q^2}{{(4\pi)}^2}\Box+m^2)\frac{\bar{D}^2}{4}\\
  (\frac{Q^2}{{(4\pi)}^2}\Box+m^2)\frac{D^2}{4} & 9\bar{\Lambda}
  \end{array}
\right)
\left(
\begin{array}{cc}
    G_{++} & G_{+-}\\
    G_{-+} & G_{--}
\end{array}
\right)
=
\left(
\begin{array}{cc}
   \delta_+ & 0\\
     0 & \delta_-
\end{array}
\right)
\end{equation}
where $\delta_+=-\frac{1}{4} \bar{D}^2\delta^8(z_1-z_2)$,
$\delta_-=-\frac{1}{4} D^2\delta^8(z_1-z_2)$.
The solution of this equation is written in the form
\begin{equation}
G=\frac{1}{81\Lambda\bar{\Lambda}-\Box{(\frac{Q^2}{16\pi^2}\Box+m^2)}^2}
\left[
\begin {array} {cc}
  9\bar{\Lambda} & -(\frac{Q^2}{16\pi^2}\Box+m^2)\frac{\bar{D}^2}{4}\\
  -(\frac{Q^2}{16\pi^2}\Box+m^2)\frac{D^2}{4} & 9\Lambda
  \end {array}
\right]
\end{equation}
   This propagator acts on columns
$$\left(
\begin{array}{c}
   \phi\\
   \bar{\phi}
\end{array}
\right)$$
where $\phi$ is a chiral superfield and $\bar{\phi}$ is an antichiral
superfield.

  The propagator in momentum representation looks like follows
\begin{eqnarray}
\label{Green}
G_{++}(k)&=&\bar{G}_1(k)\frac{-\bar{D}^2}{4}\delta_{12}\nonumber\\
G_{+-}(k)&=&G_2(k)\frac{\bar{D}^2 D^2}{16}\delta_{12}\\
G_{-+}(k)&=&G_2(k)\frac{D^2\bar{D}^2}{16}\delta_{12}\nonumber\\
G_{--}(k)&=&G_1(k)\frac{-D^2}{4}\delta_{12}\nonumber
\end{eqnarray}
 where
\begin{eqnarray}
\label{Gr2}
\delta_{12}&=&\delta^4(\theta_1-\theta_2)\nonumber\\
G_1(k)&=&\frac{9\Lambda}{k^2{(\frac{Q^2}{16\pi^2}k^2-m^2)}^2+
81\Lambda\bar{\Lambda}}\\
\bar{G}_1(k)&=&\frac{9\bar{\Lambda}}{k^2{(\frac{Q^2}{16\pi^2}k^2-m^2)}^2+
81\Lambda\bar{\Lambda}}\nonumber\\
G_2(k)&=&\frac{-\frac{Q^2}{16\pi^2}k^2+m^2}{k^2{(\frac{Q^2}{16\pi^2}k^2-m^2)}^2
+81\Lambda\bar{\Lambda}}\nonumber
\end{eqnarray}

A structure of vertices is given by the form of action (\ref{action}).
It means that we have the vertices of four types
\begin{eqnarray}
V_1&=&\xi_1 \bar{D}^{\dot{\alpha}}\bar{\sigma}
D^{\alpha}\sigma\partial_{\alpha\dot{\alpha}}(\sigma+\bar{\sigma})\nonumber\\
V_2&=&\xi_2\bar{D}_{\dot{\alpha}}\bar{\sigma}\bar{D}^{\dot{\alpha}}\bar{\sigma}
D^{\alpha}\sigma D_{\alpha}\sigma\nonumber\\
V_3&=&-\frac{m^2}{2}(e^{\sigma+\bar{\sigma}}-1-(\sigma+\bar{\sigma})-\frac{1}{2}
{(\sigma+\bar{\sigma})}^2)\\
V_4&=&\Lambda (-\frac{D^2}{4\Box})(e^{3\sigma}-1-3\sigma-\frac{9}{2}\sigma^2)+
\bar{\Lambda} (-\frac{\bar{D}^2}{4\Box})(e^{3\bar{\sigma}}-1-3\bar{\sigma}
-\frac{9}{2}\bar{\sigma}^2)\nonumber
\end{eqnarray}
The factors $(-\frac{D^2}{4\Box})$ and $(-\frac{D^2}{4\Box})$ arisen in $V_4$
since the all vertices correspond to the action written as the integrals over
whole superspace.

The eqs. (10-12) are the basis of supergraph technique allowing to develop a
perturbative consideration for the model (\ref{action}).

\vspace{5mm}
{\section{Structure of divergences}}

A structure of divergences and hence a renormalization structure are defined in
terms of superficial degree of divergences. We consider in this section a
calculation of superficial degree of divergences for the theory(\ref{action}).
The corresponding details of supergraph technique are given for example in the
book [7].

   A value of superficial degree of divergences is stipulated by the
$D$-factors
and momentum dependence of the propagator (\ref{Green}), by the derivatives in
vertices (12) and by integration over momenta in supergraphs. Let us consider
a calculation of contributions to superficial degree of divergences from
vertices, propagators and integrations.

   Each vertex of type of $V_1$ or $V_2$ contains factors $D \bar{D} \partial$
or $D \bar{D} D \bar{D}$. The canonical dimension of $D$ and $\bar{D}$ is 1/2,
the canonical dimension of $\partial_{\alpha\dot{\alpha}}$ is 1. Hence the
total
contribution from these vertices is equal to $2V_{1,2}$, where $V_{1,2}$ is
number of vertices of the type $V_{1,2}$ respectively.

Each loop corresponds to integration over 4-momentum. Hence the total
contribution from all loops is equal to $4L$, where $L$ is a number of loops.

   Each propagator $G_{--}$ or $G_{++}$ contains 2 $D$-factors (\ref{Green})
and the factor $G_1(k)$, $\bar{G}_1(k)$ (11). The two $D$-factors give the
contribution 1 and the $G_1(k), \bar{G}_1(k)$-factors give the contribution -6.
Hence the full contribution from all propagators $G_{--}$, $G_{++}$ will be
$-5L_1$ where $L_1$ is a number $G_{--}$-lines and $G_{++}$-lines in the
supergraph.

 Each propagator $G_{+-}$ or $G_{-+}$ contains 4 $D$-factors (10) and one
factor $G_2(k)$ (11). The four $D$-factors give the contribution 2 and the
$G_2(k)$-factor gives the contribution -4.Therefore the total contribution from
all propagators $G_{+-}, G_{-+}$ is equal to  $-2L_2$, where $L_2$ is a number
of $G_{+-}$-lines and $G_{-+}$lines in the supergraph.

It is known that each loop gives a contribution which is local in $\theta$-
space.
This result is based on the identity
$$\delta_{12}D^2_1\bar{D}^2_1\delta_{12}=16\delta_{12}$$
It means that four $D$-factor in each loop always can be used to create above
identity and should not be taken into account. Their total contribution $2L$
must be considered as superfluous.

   Each vertex of type $V_4$ contains two $D$-factors giving the contribution 1
and $\Box^{-1}$ giving the contribution -2. Hence the total contribution from
above vertices will be -$V_4$ where $V_4$ is a number of vertices of this type.
And, at last, the vertices of $V_3$-type do not contribute to superficial
degree
of divergences.

   As a result, the total contribution to superficial degree of divergence
$\omega$ looks like follows
\begin{equation}
\label{ind1}
 \omega=2V_{1,2}-2L_2-5L_1+2L-V_4
\end{equation}
   Certainly we should say that the eq. (\ref{ind1}) is not an exact expression
for
superficial degree of divergence but rather only an upper limit for it. The
matter is that in definite cases some of the derivatives can be transfered to
external lines of diagram and hence they give no contribution to superficial
degree of divergences. Therefore it will be more suitable to write the sign
$\leq$ in eq.(\ref{ind1}).

   The expression (\ref{ind1}) must be simplified using two useful identities.
The first one is the total number of vertices $V$ written in the form
$$V=V_{1,2}+V_3+V_4$$
and the second one is well known topological identity
$$L+V-P=1$$
where $P$ is a number of all propagators or a number of all internal lines. Two
above identities allow to rewrite the expression for $\omega$ as follows
\begin {equation}
\label{index}
\omega=2-3L_1-2V_3-3V_4
\end {equation}
This is a final result for $\omega$.

   As usual, the diagram will be divergent at $\omega\geq 0$. Let us discuss
this condition. It is evident, $\omega\geq 0$ leads to $L_1=0, V_4=0, V_3=0,1$.
Thus, the divergent diagrams cannot contain the $V_4$-type vertices and
$G_{++}$-, $G_{--}$-type lines. They can include no more that one vertex of
$V_3$-type. We see a number of divergent structures is more restricted in
compare with non-supersymmetric case [18]. In particular we have some sort of
non-renormalization theorem: the vertex of $V_4$-type is always finite.

  The condition (\ref{index}) shows that a divergent diagram can include an
arbitrary number of $G_{+-}$-, $G_{-+}$-lines and arbitrary number of
$V_{1,2}$-
type vertices. But if, for example, $\xi_1=\xi_2=0$ then the vertices of
$V_{1,2}$-type are absent at all. The only vertices which can present in
divergent diagrams are now the $V_3$-type ones. But their number is exactly
equal to 1 and corresponds only to one-loop diagrams. Thus at $\xi_1=\xi_2=0$
the theory becomes to be super-renormalizable. We will show that in infrared
limit the effective couplings $\xi_1(t)$ and $\xi_2(t)$ tend to zero and namely
above case is realized.

\vspace{5mm}
{\section{One-loop divergences}}

 We will investigate the one-loop divergences in the theory under consideration
in framework of dimensional regularization.

The action of the theory written formally (in sense of dimensional reduction)
in $d$ dimensions has the form:
\begin {eqnarray}
S&=&\mu^{-\epsilon}\big(\int d^d x d^4\theta
    (-\frac{Q^2}{2{(4\pi)}^2}\bar{\sigma}\Box\sigma+\bar{D}^{\dot\alpha}
\bar{\sigma}D^{\alpha}\sigma\times\nonumber\\
&\times&(\xi_1{\partial_{\alpha}}_{\dot{\alpha}}(\sigma+\bar{\sigma})
+\xi_2\bar{D}_{\dot{\alpha}}\bar{\sigma} D_{\alpha}\sigma)-
\frac{m^2}{2} e^{\sigma+\bar{\sigma}})+(\Lambda\int d^d x d^2\theta
e^{3\sigma}+
h.c.)\big)
\end {eqnarray}

Let us begin with one-loop supergraph providing a calculation of
counterterms leading to renormalization of the parameters $\xi_1$, $\xi_2$.
The corresponding supergraphs are given by the  Fig.1 and Fig.2.
\begin{center}
\begin{picture}(100,100)
\put(50,50){\circle{40}}
\put(30,50){\line(-1,-1){20}}
\put(15,40){\line(0,-1){10}}
\put(10,20){$D^{\alpha}$}
\put(35,70){\line(0,-1){10}}
\put(30,70){$\bar{D}_{\dot{\alpha}}$}
\put(30,50){\line(-1,1){20}}
\put(15,70){\line(0,-1){10}}
\put(35,40){\line(0,-1){10}}
\put(30,20){$D_{\alpha}$}
\put(10,75){$\bar{D}^{\dot{\alpha}}$}
\put(70,50){\line(1,0){20}}
\put(80,40){$\partial_{\beta\dot{\beta}}$}
\put(80,55){\line(0,-1){10}}
\put(65,40){\line(0,-1){10}}
\put(65,70){\line(0,-1){10}}
\put(70,75){$D^{\beta}$}
\put(70,20){$\bar{D}^{\dot{\beta}}$}
\put(45,80){$G_{+-}$}
\put(45,20){$G_{-+}$}
\put(40,0){Fig.1}
\end{picture}
\end{center}
\begin{center}
\begin{picture}(100,100)
\put(50,50){\circle{40}}
\put(30,50){\line(-1,-1){20}}
\put(30,50){\line(-1,1){20}}
\put(70,50){\line(1,-1){20}}
\put(70,50){\line(1,1){20}}
\put(15,40){\line(0,-1){10}}
\put(10,20){$D^{\alpha}$}
\put(15,70){\line(0,-1){10}}
\put(10,75){$\bar{D}^{\dot{\alpha}}$}
\put(85,70){\line(0,-1){10}}
\put(85,40){\line(0,-1){10}}
\put(85,20){$D^{\beta}$}
\put(85,75){$\bar{D}^{\dot{\beta}}$}
\put(35,40){\line(0,-1){10}}
\put(30,20){$D_{\alpha}$}
\put(35,70){\line(0,-1){10}}
\put(30,70){$\bar{D}_{\dot{\alpha}}$}
\put(65,40){\line(0,-1){10}}
\put(65,70){\line(0,-1){10}}
\put(70,75){$D_{\beta}$}
\put(70,20){$\bar{D}_{\dot{\beta}}$}
\put(45,80){$G_{+-}$}
\put(45,20){$G_{-+}$}
\put(40,0){Fig.2}
\end{picture}
\end{center}
The supergraphs associated with Fig.1 and Fig.2 lead to the following
contributions respectively
\begin{eqnarray}
{S_1}^{(1)}&=&72\mu^{-\epsilon}\xi_1\xi_2\int d^4\theta_1 d^4\theta_2
\frac{d^d p_1 d^d p_2}{{(2\pi)}^{2d}}\times\nonumber\\
&\times&\bar{D}^{\dot{\alpha}}\bar{\sigma}(-p_1,\theta_1)
D^{\alpha}\sigma(-p_2,\theta_1)
\partial_{\beta\dot{\beta}}(\sigma(p,\theta_2)+ \bar{\sigma}(p,\theta_2))
\times\nonumber\\
&\times&\int\frac{d^d k}{{(2\pi)}^d} D_{\alpha}\bar{D}^{\dot{\beta}}G_{+-}(k)
\bar{D}_{\dot{\alpha}}D^{\beta} G_{-+}(k+p)\\
{S_2}^{(1)}&=&72\mu^{-\epsilon}\xi_2^2\int d^4\theta_1 d^4\theta_2
\frac{d^d p_1 d^d p_2 d^d p_3}{{(2\pi)}^{3d}}\times\nonumber\\
&\times&\bar{D}^{\dot{\alpha}}\bar{\sigma}(-p_1,\theta_1)
D^{\alpha}\sigma(-p_2,\theta_1)
\bar{D}_{\dot{\beta}}\bar{\sigma}(p_3,\theta_2)
 D_{\beta}\sigma(p-p_3,\theta_2)\times\nonumber\\
&\times&\int\frac{d^d k}{{(2\pi)}^d}
D_{\alpha}\bar{D}^{\dot{\beta}}G_{+-}(k) \bar{D}_{\dot{\alpha}} D^{\beta}
G_{-+}(k+p)\nonumber
\end{eqnarray}
Here ${S_1}^{(1)}$ and ${S_2}^{(1)}$ are one-loop divergent corrections to
vertices $V_1$ and $V_2$ correspondingly, $p_1$, $p_2$ and $p_3$ are external
momenta, $p=p_1+p_2$. $\mu$ is a standard arbitrary parameter of mass dimension
introduced in dimensional regularization and $\epsilon=4-d$. Using the explicit
form of propagators (10,11) one can write

 \begin{eqnarray}
 \label{loop}
{S_1}^{(1)}&=&72\mu^{-\epsilon}\xi_1\xi_2\int d^4\theta_1 d^4\theta_2
\frac{d^d p_1 d^d p_2}{{(2\pi)}^{2d}}\times\nonumber\\
&\times&\bar{D}^{\dot{\alpha}}\bar{\sigma}(-p_1,\theta_1)
D^{\alpha}\sigma(-p_2,\theta_1)
\partial_{\beta\dot{\beta}}(\sigma(p,\theta_2)+ \bar{\sigma}(p,\theta_2))
\times\nonumber\\
&\times& \int\frac{d^d k}{{(2\pi)}^d} \frac{\bar{D}_{1\dot{\alpha}}
D^{\beta}_2 \bar{D}^2_2 D^2_1}{16}\delta_{12}\times\\
&\times&\frac{D_{1\alpha}\bar{D}^{\dot{\beta}}_2 \bar{D}^2_1 D^2_2}{16}
\delta_{12}G_2(k)G_2(k+p)\nonumber\\
\nonumber\\
{S_2}^{(1)}&=&72\mu^{-\epsilon}\xi_2^2\int d^4\theta_1 d^4\theta_2
\frac{d^d p_1 d^d p_2 d^d p_3}{{(2\pi)}^{3d}}\times\nonumber\\
&\times&\bar{D}^{\dot{\alpha}}\bar{\sigma}(-p_1,\theta_1)
D^{\alpha}\sigma(-p_2,\theta_1)
\bar{D}_{\dot{\beta}}\bar{\sigma}(p_3,\theta_2)
 D_{\beta}\sigma(p-p_3,\theta_2)
\times\nonumber\\
&\times&\frac{d^d k}{{(2\pi)}^d}
\frac{\bar{D}_{1\dot{\alpha}}D^{\beta}_2 D^2_1
\bar{D}^2_2}{16}\delta_{12}\times\nonumber\\
&\times&\frac{D_{1\alpha}
\bar{D}^{\dot{\beta}}_2 \bar{D}^2_1 D^2_2}{16}\delta_{12}
G_2(k)G_2(k+p)\nonumber
\end {eqnarray}

 Let us start to simplify the eqs (\ref{loop}).
  Using the known properties of the covariant derivatives one can write
\begin{eqnarray}
& &\frac{\bar{D}_1^{\dot{\alpha}}D^{\beta}_2 D^2_1
\bar{D}^2_2}{16}\delta_{12}
\frac{D_1^{\alpha}\bar{D}^{\dot{\beta}}_2 \bar{D}^2_1 D^2_2}{16}
\delta_{12}=\nonumber\\
&=&\frac{\bar{D}^2_2 D^2_1 D^{\beta}_2 \bar{D}^{\dot{\beta}}_2}{16}\delta_{12}
\frac{\bar{D}_1^{\dot{\alpha}} D_1^{\alpha}\bar{D}^2_1 D^2_2}{16}\delta_{12}=
\frac{\bar{D}^2_1 D^2_2}{16}\partial_2^{\beta\dot{\beta}}\delta_{12}
\partial_1^{\alpha\dot{\alpha}}\frac{D^2_1 \bar{D}^2_2}{16}\delta_{12}
\nonumber
\end{eqnarray}
where
$k_{\dot{\alpha}\beta}=2\bar{\sigma}_{\dot{\alpha}\beta}^m k_m$.
   Therefore
\begin{eqnarray}
{S_1}^{(1)}&=&72\mu^{-\epsilon}\xi_1\xi_2\int d^4\theta_1 d^4\theta_2
\frac{d^d p_1 d^d p_2}{{(2\pi)}^{2d}}\times\nonumber\\
&\times&\bar{D}^{\dot{\alpha}}
\bar{\sigma}(-p_1,\theta_1)D^{\alpha}\sigma(-p_2,\theta_1)
\partial_{\beta\dot{\beta}}(\sigma(p,\theta_2)+ \bar{\sigma}(p,\theta_2))
\times\nonumber\\
&\times&\int\frac{d^d k}{{(2\pi)}^d}\frac{D^2\bar{D}^2}{16}\delta_{12}
\frac{\bar{D}^2 D^2}{16}\delta_{12}\times\nonumber\\
&\times&4\sigma^m_{\alpha\dot{\alpha}}\bar{\sigma}^{n\beta\dot{\beta}}
k_m {(k+p)}_n G_2(k)G_2(k+p)\\
{S_2}^{(1)}&=&72\mu^{-\epsilon}\xi_2^2\int d^4\theta_1 d^4\theta_2
\frac{d^d p_1 d^d p_2 d^d p_3}{{(2\pi)}^{3d}}\times\nonumber\\
&\times&\bar{D}^{\dot{\alpha}}\bar{\sigma}(-p_1,\theta_1)
D^{\alpha}\sigma(-p_2,\theta_1)
\bar{D}_{\dot{\beta}}\bar{\sigma}(p_3,\theta_2)
D_{\beta}\sigma(p-p_3,\theta_2)
\times\nonumber\\
&\times& \int\frac{d^d k}{{(2\pi)}^d}\frac{D^2\bar{D}^2}{16}
\delta_{12} \frac{\bar{D}^2 D^2}{16}\delta_{12}
4\sigma^m_{\alpha\dot{\alpha}}\bar{\sigma}^{n\beta\dot{\beta}}
k_m{(k+p)}_n G_2(k)G_2(k+p)\nonumber
\end{eqnarray}

Now one transfers the covariant derivatives from one delta-function to another
using integration by parts. Note the divergence can arise only when all
covariant derivatives act on one delta-function. As a result we obtain the
expression

$$\delta_{12}\frac{\bar{D}^2 D^2\bar{D}^2 D^2}{256}\delta_{12}=
-k^2\delta_{12}\frac{\bar{D}^2 D^2}{16}\delta_{12}=-k^2\delta_{12}$$
After these transformations, the corrections to vertices $V_1$, $V_2$ take the
form

\begin{eqnarray}
\label{cor1}
S_1^{(1)}&=&-288\mu^{-\epsilon}\xi_1\xi_2\int d^4\theta
\int\frac{d^d p_1 d^d p_2}{{(2\pi)}^{2d}}\times\nonumber\\
&\times&\bar{D}^{\dot{\alpha}}\bar{\sigma}(-p_1,\theta)
D^{\alpha}\sigma(-p_2,\theta)
\partial_{\beta\dot{\beta}}(\sigma(p,\theta)+ \bar{\sigma}(p,\theta))
\sigma^m_{\alpha\dot{\alpha}}\bar{\sigma}^{n\beta\dot{\beta}}I_{mn}
\nonumber\\
S_2^{(1)}&=&-288\mu^{-\epsilon}\xi_2^2\int d^4\theta
\int\frac{d^d p_1 d^d p_2 d^d p_3}{{(2\pi)}^{3d}}\times\nonumber\\
&\times&\bar{D}^{\dot{\alpha}}\bar{\sigma}(-p_1,\theta)
D^{\alpha}\sigma(-p_2,\theta)
\bar{D}_{\dot{\beta}}\bar{\sigma}(p_3,\theta)
 D_{\beta}\sigma(p-p_3,\theta)
\sigma^m_{\alpha\dot{\alpha}}\bar{\sigma}^{n\beta\dot{\beta}}I_{mn}
\nonumber
\end{eqnarray}
where

\begin {equation}
I_{mn}=\int \frac{d^d k}{{(2\pi)}^d} k^2 k_m{(k+p)}_n G_2(k)G_2(k+p)
\end {equation}

 Let us investigate this integral in details. We present the Green function
$G_2(k)$ (\ref{Gr2}) in the form
$$G_2(k)=-\frac{1}{A^2}\frac{k^2-\frac{m^2}{A^2}} {(k^2+a_1)(k^2+a_2)(k^2+a_3)}
$$
where all quantities $a_1$, $a_2$, $a_3$ are different and
$A^2=\frac{Q^2}{16\pi^2}$.
The function $G_2(k)$ is symmetric with respect to $a_1$, $a_2$, $a_3$. Taking
into account this property we can write
\begin{equation}
\label{div}
G_2(k)=-\frac{1}{A^2(k^2+a_2)(k^2+a_3)}+\frac{1}{A^2}\frac{\frac{m^2}{A^2}+a_1}
{(k^2+a_1)(k^2+a_2)(k^2+a_3)}
\end{equation}
   Simple power counting allows to make a conclusion that only first term of
eq.(\ref{div}) gives contribution to divergent part of the integral $I_{mn}$.
It means that the divergent part $I_{mn}$ coincides with divergent part of the
following integral
\begin{equation}
I'_{mn}=\frac{1}{A^4} \int \frac{d^d k}{{(2\pi)}^d}\frac{k^2 k_m k_n}
{(k^2+a_2)(k^2+a_3)({(k+p)}^2+a_2)({(k+p)}^2+a_3)}
\end{equation}
   Using the analogous transformation one can rewrite this integral in the form
\begin{equation}
I'_{mn}=\frac{1}{A^4}(\int \frac{d^d k}{{(2\pi)}^d}\frac{k_m k_n}
{(k^2+a_3)({(k+p)}^2+a_2)({(k+p)}^2+a_3)}+fin)
\end{equation}
where $fin$ denotes finite part of the integral.
   Now we use the Feynman representation and transform the divergent part of
the integral $I'_{mn}$ to the following expression
\begin{equation}
I''_{mn}=\frac{2}{A^4}\int \frac{d^d k}{{(2\pi)}^d}
\frac{k_m k_n dx dy dz \delta(1-x-y-z)}
{{(k^2+ 2kp(x+y)+a_2 x+a_3(y+z))}^3}
\end{equation}

   Taking into account the known formula

\begin{eqnarray}
\int \frac{d^d k}{{(2\pi)}^d}\frac{k_m k_n}{{(k^2+M^2+2kq)}^N}&=&
\frac{1}{{(4\pi)}^{d/2}\Gamma(N)}\Big\{\frac{q_m q_n \Gamma(N-d/2)}
{{(M^2-q^2)}^{N-d/2}}+\\
&+&\frac{1}{2}\eta_{mn}\frac{\Gamma(N-d/2-1)}{{(M^2-q^2)}^{N-d/2-1}}\Big\}
\nonumber
\end{eqnarray}
and that $N=3$, $d=4-\epsilon$ we obtain the divergent part of the integral
$I''_{mn}$ as follows
\begin{equation}
J_{mn}=\frac{1}{16\pi^2}\frac{1}{A^4}\eta_{nm}\Gamma(\frac{\epsilon}{2})
\int\frac{dx dy dz \delta(1-x-y-z)}
{{(a_2x+a_3(y+z)-p^2{(x+y)}^2)}^{\epsilon/2}}
\end{equation}

 To extract the pole part we use
$\Gamma(\frac{\epsilon}{2})=\frac{2}{\epsilon}+fin$,
where $fin$ is finite part at $\epsilon\rightarrow 0$. As a result one obtains

\begin {equation}
\label{J}
{J_{mn}}_{div}=\frac{2}{16\pi^2\epsilon A^4}\eta_{nm}
\end {equation}
The eq. (\ref{J}) allows to write one-loop quantum corrections (\ref{loop})
in the form
\begin{eqnarray}
\label{correct}
S_1^{(1)}&=&-576\mu^{-\epsilon}\frac{\xi_1\xi_2}{A^4}\int d^4\theta d^4 x
\times\nonumber\\
&\times&\bar{D}^{\dot{\alpha}}\bar{\sigma}(x,\theta)D^{\alpha}\sigma(x,\theta)
\partial_{\alpha\dot{\alpha}}(\sigma(x,\theta)+ \bar{\sigma}(x,\theta))
(\frac{2}{16\pi^2\epsilon}+fin)\equiv\\
&\equiv& S^{(1)}_{1_{div}}+S^{(1)}_{1_{fin}}\nonumber\\
S_2^{(1)}&=&-576\mu^{-\epsilon}\frac{\xi^2_2}{A^4}\int d^4\theta d^4 x
\times\nonumber\\
&\times&\bar{D}^{\dot{\alpha}}\bar{\sigma}(x,\theta)
D^{\alpha}\sigma(x,\theta)
\bar{D}_{\dot{\alpha}}\bar{\sigma}(x,\theta)
 D_{\alpha}\sigma(x,\theta)
(\frac{2}{16\pi^2\epsilon}+fin)\equiv\nonumber\\
&\equiv& S^{(1)}_{2_{div}}+S^{(1)}_{2_{fin}}\nonumber
\end{eqnarray}

In order to renormalize the theory we introduce the one-loop counterterms\\
$-{S_1^{(1)}}_{div}$, $-{S_2^{(1)}}_{div}$. It corresponds to the following
renormalization transformation

\begin{eqnarray}
\label{ren0}
Q^2_{(0)}=\mu^{-\epsilon}Z_Q Q^2\nonumber\\
\xi_{1(0)}=\mu^{-\epsilon}Z_1\xi_1\\
\xi_{2(0)}=\mu^{-\epsilon}Z_2\xi_2\nonumber
\end{eqnarray}
where
$Q^2_{(0)}, \xi_{1(0)}, \xi_{2(0)}$ are the bare parameters and
$Q^2, \xi_1, \xi_2 $- are the renormalized ones. As a result one obtains

\begin{equation}
\label{ren12}
Z_1=Z_2=(1+\frac{72\xi_2}{\pi^2 A^4\epsilon})
\end{equation}
We see that in one-loop approximation there is the same independent
renormalization constant both for $\xi_1$ and $\xi_2$. It means in particular,
if we put $\xi_2^{(0)}=c\xi_1^{(0)}$ where $c$ is a constant, then the
renormalized parameters $\xi_1$ and $\xi_2$ will satisfy the same relation
$\xi_2=c\xi_1$. One-loop renormalization does not destroy the relationship
between the parameters.

Next step is a calculation of $Z_Q$. Let us consider the supergraph given on
Fig.3
\begin{center}
\begin{picture}(100,100)
\put(50,50){\circle{40}}
\put(30,50){\line(-1,0){20}}
\put(70,50){\line(1,0){20}}
\put(15,55){\line(0,-1){10}}
\put(80,55){\line(0,-1){10}}
\put(80,40){$\partial_{\alpha\dot{\alpha}}$}
\put(10,60){$\partial_{\beta\dot{\beta}}$}
\put(65,40){\line(0,-1){10}}
\put(65,70){\line(0,-1){10}}
\put(70,75){$D^{\alpha}$}
\put(70,20){$\bar{D}^{\dot{\alpha}}$}
\put(35,40){\line(0,-1){10}}
\put(30,20){$D^{\beta}$}
\put(35,70){\line(0,-1){10}}
\put(30,70){$\bar{D}^{\dot{\beta}}$}
\put(45,80){$G_{+-}$}
\put(45,20){$G_{-+}$}
\put(40,0){Fig.3}
\end{picture}
\end{center}
The corresponding contribution looks like this
\begin{eqnarray}
S_Q^{(1)}&=&-18\mu^{-\epsilon}\xi^2_1\int d^4\theta_1 d^4\theta_2\int \frac
{d^d p}{{(2\pi)}^d}
\partial_{\beta\dot{\beta}}\bar{\sigma}(-p,\theta_1)
\partial_{\alpha\dot{\alpha}}\sigma(p,\theta_2)\times\\
&\times&\int\frac{d^d k}{{(2\pi)}^d}
\frac{\bar{D}^{\dot{\beta}} D^{\alpha}\bar{D}^2 D^2}{16}\delta_{12}
\frac{D^{\beta}\bar{D}^{\dot{\alpha}} D^2 \bar{D}^2}{16}\delta_{12}
G_2(k)G_2(k+p)\nonumber
\end{eqnarray}
 Carrying out the transformations analogous to used above we obtain
\begin{equation}
S_Q^{(1)}=-\mu^{-\epsilon}\int d^4\theta d^d x\partial_{\alpha\dot{\alpha}}
\sigma\partial^{\alpha\dot{\alpha}}\bar{\sigma}
(\frac{18\xi_1^2{(4\pi)}^4}{Q^4\epsilon}+fin)\equiv S^{(1)}_{Q_{div}}+
S^{(1)}_{Q_{fin}}
\end{equation}
After introducing the one-loop counterterm $-{S_Q^{(1)}}_{div}$ one will obtain
using
(\ref{ren0})
\begin {equation}
\label{Z}
Z_Q=(1+\frac{32\pi^2}{Q^2}\frac{18\xi_1^2{(4\pi)}^4}{Q^4\epsilon})
\end {equation}

   So we have studied renormalization of $\xi_1$, $\xi_2$ and $Q^2$. As for
$\Lambda$, it was already noted that all diagrams containing vertex of type
$V_4$ are finite, it means that the coupling $\Lambda$ is not renormalized.

   Now it remains to investigate renormalization of $m^2$. It follows from
(\ref{index}), that divergent diagrams can contain no more than one vertex type
of $V_3$ corresponding to coupling constant $m^2$. All other possible vertices
should be of types of $V_1$ or $V_2$.

We will study the divergent corrections to $m^2$ in the case when $\xi_1$=
$\xi_2=0$.It means that the vertices $V_1$ and $V_2$ are absent at all. It will
be proved further that this case corresponds to infrared limit of the theory.
It means that only $V_3$-type vertex can be presented in the diagrams giving
contribution to divergent correction to $m^2$. All these diagrams contain only
one vertex $V_3$-type, one internal line $G_{+-}$-type and the arbitrary
number of external lines corresponding to $\sigma$, $\bar{\sigma}$
\begin{center}
\begin{picture}(100,100)
\put(50,50){\circle{40}}
\put(30,50){\line(-1,-1){20}}
\put(30,50){\line(-1,1){20}}
\put(20,45){$\vdots$}
\put(75,50){$G_{+-}$}
\put(40,0){Fig.4}
\end{picture}
\end{center}

   Let us consider such a diagram with given number $N$ of external lines,
$l$ from those are chiral and other are antichiral. Contribution of
this diagram has the form
\begin{equation}
-\frac{m^2}{2}\int d^8 z\frac{\sigma^l(z)\bar{\sigma}^{N-l}(z)}{l!(N-l)!}
 G_{+-}(z,z)
\end{equation}
  Sum of all these contributions is equal to
\begin{equation}
S_3=-\frac{m^2}{2}\int d^8 z \sum_{N=0}^{\infty}\sum_{l=0}^{N}
\frac{\sigma^l(z)\bar{\sigma}^{N-l}(z)}{l!(N-l)!}G_{+-}(z,z)=
\frac{m^2}{2}\int d^8 z e^{\sigma+\bar{\sigma}} G_{+-}(z,z)
\end{equation}
In momentum representation the $S_3$ can be written as follows
\begin{equation}
S_3=-\frac{m^2}{2}\int d^4\theta d^d x e^{\sigma+\bar{\sigma}}\int
\frac{d^d k}{{(2\pi)}^d}\frac{-A^2 k^2+m^2}
{k^2{(-A^2 k^2+m^2)}^2- 81\Lambda\bar{\Lambda}}
\frac{\bar{D}^2_1 D^2_1}{16}\delta_{11}
\end{equation}
Taking into account
$\frac{\bar{D}^2_1 D^2_1}{16}\delta_{11}=\frac{\bar{D}^2_1 D^2_1}{16}
 \delta_{12}\big|_{\theta_1=\theta_2}=1 $ we get
 \begin{equation}
S_3=-\frac{m^2}{2}\int d^4\theta  d^d x e^{\sigma+\bar{\sigma}}
\int\frac{d^d k}{{(2\pi)}^d}
\frac{-A^2 k^2+m^2}{k^2{(-A^2 k^2+ m^2)}^2-81\Lambda\bar{\Lambda}}
\end{equation}

Let us consider the integral
$$\int \frac{d^d k}{{(2\pi)}^d}
\frac{-A^2 k^2+m^2}{k^2{(-A^2 k^2+m^2)}^2-81\Lambda\bar{\Lambda}}$$
A divergent part of this integral coincides with the divergent part of
the following integral
$$-\frac{1}{A^2}\int \frac{d^d k}{{(2\pi)}^d}\frac{1}{(k^2+a)(k^2+b)}$$
where $a$ and $b$ are the roots of denominator of initial integral.
The last integral is equal to $-\frac{1}{8\pi^2\epsilon A^2}$. As a result one
obtains
$$ S_3=\frac{m^2}{2}\int d^d x d^4\theta e^{\sigma+\bar{\sigma}}
(\frac{2}{16\pi^2 A^2\epsilon}+fin)$$

To cancel the divergence we should introduce a counterterm $-{S_3}_{div}$. It
corresponds to mass renormalization
\begin{eqnarray}
\label{rmass}
m^2_0&=&\mu^{-\epsilon} Z_m m^2\nonumber\\
Z_{m^2}&=&1+\frac{2}{16\pi^2 A^2\epsilon}=1+\frac{2}{Q^2\epsilon}
\end{eqnarray}
Here $m^2_0$ is a bare mass and $m^2$ is a renormalized one.
The eqs (28,29,32,37) define the one-loop renormalization of the theory under
consideration.

\vspace{5mm}
{\section{Running couplings and infrared freedom}}

  In this section we will analyse renormalization group equations and
investigate asymptotical behaviour of running couplings.

Let us start with eqs (28,29,32). These equations lead to the following beta-
functions
\begin{eqnarray}
\label{rgr}
\beta_{\xi_1}=\frac{72{(16\pi)}^2}{Q^4}\xi_1\xi_2\nonumber\\
\beta_{\xi_2}=\frac{72{(16\pi)}^2}{Q^4}\xi_2^2\nonumber\\
\beta_{Q^2}=32\pi^2\frac{18{(16\pi)}^2 \xi^2_1}{Q^4}\nonumber
\end{eqnarray}

As a result the equations for running couplings have the form
\begin{eqnarray}
\label{eqr}
\frac{d\xi_1}{dt}=a\frac{\xi_1\xi_2}{Q^4}\nonumber\\
\frac{d\xi_2}{dt}=a\frac{\xi_2^2}{Q^4}\\
\frac{d Q^2}{dt}=b\frac{\xi_1^2}{Q^4}\nonumber
\end{eqnarray}
where $a=2^{11}3^2\pi^2$, $b=3^2 2^{14}\pi^4$.
The solutions of these equations look like this
\begin{eqnarray}
\xi_1(t)&=&\frac{\xi_1}{\xi_2}\xi_2(t)\nonumber\\
Q^2(t)&=& Q^2+8\pi^2\frac{\xi_1^2}{\xi_2^2}(\xi_2(t)-\xi_2)\nonumber\\
t&=&\frac{1}{2^{11} 3^2\pi^2}\Big\{-[Q^2-8\pi^2\frac{\xi^2_1}{\xi^2_2}]
(\frac{1}{\xi_2(t)}-\frac{1}{\xi_2})-\nonumber\\
&-&16\pi^2{(\frac{\xi_1}{\xi_2})}^2 [Q^2-8\pi^2\frac{\xi^2_1}{\xi^2_2}]
\ln\frac{\xi_2(t)}{\xi_2}+\\
&+&64\pi^4{(\frac{\xi_1}{\xi_2})}^4(\xi_2(t)-\xi_2)\Big\}\nonumber
\end{eqnarray}

Let us investigate a behaviour of running couplings $\xi_1(t)$, $\xi_2(t)$
and $Q^2(t)$ in infrared domain when $t\rightarrow-\infty$. It is easy to see
that in this case $\xi_2(t)\rightarrow 0$ and hence $\xi_1(t)\rightarrow 0$.
It means that $\xi_1^{(0)}$= $\xi_2^{(0)}$= 0 is an infrared fixed point. For
$Q^2(t)$ we obtain $Q^2(t)\rightarrow Q^2-8\pi^2\frac{\xi^2_1}{\xi_2}$.
If we take quantities of initial $\xi_1$ and $\xi_2$ so that they correspond to
infrared fixed point $\xi_1=\xi_2=0$ one gets $Q^2(t)\rightarrow Q^2$.
In particular, only the diagrams given on Fig.4 can contribute to mass
renormalization in infrared limit.

To investigate a behaviour of running mass we should use a notion of scaling
dimension of superfields. We note that the action of the theory (\ref{action})
is invariant under the transformations

\begin{eqnarray}
\label{trans}
\delta\sigma=(x^a \partial_a +\frac{1}{2}\theta^{\alpha} D_{\alpha})\sigma+1\\
\delta\bar{\sigma}=(x^a
\partial_a+\frac{1}{2}\bar{\theta}_{\dot{\alpha}}\bar{D}
^{\dot{\alpha}})\bar{\sigma}+1\nonumber
\end{eqnarray}
   Let $V$ is some function depending on superfields $\sigma$, $\bar{\sigma}$
and their derivatives $\partial_a\sigma$, $\partial_a\bar{\sigma}$,
$D_{\alpha}\sigma$, $\bar{D}_{\dot{\alpha}}\bar{\sigma}$,\ldots. We call the
$V$
has the scaling dimension $\Delta$ if the transformation law of $V$ under
transformations (\ref{trans}) looks like this
\begin{equation}
\delta V[\sigma,\bar{\sigma}]=(x^a\partial_a+
\frac{1}{2}\theta^{\alpha} D_{\alpha}+
\frac{1}{2}\bar{\theta}_{\dot{\alpha}} \bar{D}^{\dot{\alpha}}+\Delta) V
\end{equation}

 It is easy to see that the superfields $\sigma$, $\bar{\sigma}$ have no
definite scaling dimension, the derivatives $\partial_a\sigma$,
$\partial_a\bar{\sigma}$ have scaling dimension equal to 1, the spinor
derivatives $D_{\alpha}\sigma$,$\bar{D}_{\dot{\alpha}}\bar{\sigma}$ have no
definite scaling dimensions. However, the functions $e^{\sigma}$,
$e^{\bar{\sigma}}$ have definite scaling dimensions $\Delta$=1.

  Let us fulfil the transformations $\sigma\rightarrow\alpha\sigma$,
$\bar{\sigma}\rightarrow\alpha\bar{\sigma}$ and
$S\rightarrow\frac{1}{\alpha^2}S$
in the action (\ref{action}) at $\xi_1=\xi_2=0$. It leads to the following
action depending on arbitrary real parameter $\alpha$
\begin{eqnarray}
\label{iract}
S=\int d^8 z(-\frac{1}{2}\frac{Q^2}{16\pi^2}\bar{\sigma}\Box\sigma-
\frac{m^2}{2\alpha^2}e^{\alpha(\sigma+\bar{\sigma})})+\\
+(\frac{\Lambda}{\alpha^2}\int d^6 z e^{3\alpha\sigma}+ h.c.)\nonumber
\end{eqnarray}

  We consider a calculation of renormalization constant $Z_m$ in the theory
(\ref{iract}). The only modification in comparison with eq. (\ref{rmass}) is
that we should use the propagator $\alpha^2G_{+-}$ in supergraph given by
Fig.4. The parameter $\alpha$ is resulted here because of expansion of
$e^{\alpha(\sigma+\bar{\sigma})}$. It leads immediately to
$$ Z_{m^2}=1+\frac{2\alpha^2}{Q^2\epsilon}$$
Therefore the equation for running mass will be
\begin{equation}
\label{eqren}
\frac{d m^2(t)}{dt}=\frac{2\alpha^2 m^2(t)}{Q^2}+d_{m^2}m^2(t)
\end{equation}
where $\Delta_{m^2}$ is a scaling dimension of  $m^2(t)$.

To find $\Delta_{m^2}$ we consider the term
$\frac{m^2}{2\alpha^2}e^{\alpha(\sigma+\bar{\sigma})}$ in the action
(\ref{iract}).
The scaling dimension of this term is -2, $\alpha$ is dimensionless and scaling
dimension of $e^{\alpha(\sigma+\bar{\sigma})}$ is $2\alpha$. Hence
$\Delta_{m^2}=2-2\alpha$.
Therefore the equation (\ref{eqren}) looks like this
\begin{eqnarray}
\frac{d m^2(t)}{dt}&=&(2-2\alpha+\frac{2\alpha^2}{Q^2})m^2(t)\\
m^2(0)&=&m^2\nonumber
\end{eqnarray}
where we took into account that $Q^2(t)= Q^2$ in infrared limit. A solution
of this equation can be written in the form
\begin{equation}
\label{efm}
m^2(t)=m^2\exp((2-2\alpha+\frac{2\alpha^2}{Q^2})t)
\end{equation}
It is evident that at $2-2\alpha+\frac{2\alpha^2}{Q^2}>0$
we get $m^2(t)\rightarrow 0$ in infrared limit. It corresponds to
$\kappa^2(t)\rightarrow \infty$ where $\kappa^2(t)$ is running gravitational
constant.

  As for coupling constant $\Lambda$, its beta-function is equal to zero since
the vertex of $V_4$-type is always finite (see section 3) and the fields
$\sigma$, $\bar{\sigma}$ are not renormalized. It means, in particular, that we
can put $\Lambda=\bar{\Lambda}=0$ in the action (\ref{action}) or (\ref{iract})
without breakdown of renormalizability of the theory.

\vspace{5mm}
{\section{Summary}}

  We have formulated a new model of chiral $N=1$ superfield in $D=4$ flat
superspace. This model is generated by superconformal anomaly of matter
superfields and can be considered as a simplified model of quantum supergravity
in low-energy domain. The features of the model are its complete superfield
formulation, non-trivial interactions of chiral and antichiral superfields,
presence of five couplings, the three of them are dimensionless, and higher
(four) derivatives in a kinetic term. The model is a natural supersymmetric
generalization of the model of low-energy quantum gravity given by Antoniadis
and Mottola \cite{AM}.

The analysis of superficial degree of divergence shows that the model under
consideration leads to decrease of number of divergent structures in comparison
with non-supersymmetric model \cite{AM}. We have calculated one-loop
counterterms and investigated equations for effective couplings. It is shown
that this model is infrared free and moreover it is superrenormalizable in
infrared limit.

An interesting feature of the model is non-renormalization theorem according
to which the vertex $\Lambda\int d^6z e^{3\sigma}$ has no divergent
corrections.
Since the superfield $\sigma$ is not renormalized in the model we  get that the
parameter $\Lambda$ (cosmological constant) is always finite. The analogous
vertex including cosmological constant in corresponding non-supersymmetric
model \cite{AM} gets divergent corrections. As a result, unlike of
non-supersymmetric case, the beta-function for $\Lambda$ is equal to zero in
our model. It means the mechanism leading to vanishing of effective
cosmological
constant given in Ref.\cite{AM} will work only if a supersymmetry is violated.

The suggested model unlike of full superfield supergravity has simple enough
superfield structure and can be applied for consideration of various aspects of
quantum supergravity in infrared domain.

We should like to pay attention that our model has been formulated within
$N=1$, $D=4$ superfield supergravity. It is well known that there are other
so-called non-minimal or new minimal versions of the supergravity (see f.e.
[7]). It would be interesting to develop an approach analogous to one under
consideration for those versions.\vspace{5mm}

{\Large\bf{Acknowledgements}}

The authors are grateful to S.M.Kuzenko and S.D.Odintsov for discussions of
some aspects of the work. The paper has been finished during a visit of I.L.B.
to Institute of Physics at Humboldt Berlin University. He would like to express
the gratitude to D.Lust, D.Ebert, H.Dorn, G.Cardoso, C.Preitschopf, M.Schmidt
and C.Schubert for their hospitality, support and interesting discussions.
The visit was supported by DFG under contract DFG-436 RUS 113. The work was
supported in part by ISF under the grant No. RI1300 and by RFBR under the
project No. 94-02-03234.
\newpage
\vspace{5mm}
{\large\bf{Appendix}}
\vspace{2mm}

   We consider here the component form of the action (6).
Let us write the component decomposition of chiral scalar superfield $\sigma$
\setcounter{equation}{0}
\renewcommand{\theequation}{\Roman{equation}}
\begin{equation}
\sigma(x,\theta,\bar{\theta})=(1+\theta^{\alpha}
{\sigma^m}_{\alpha\dot{\alpha}}\bar{\theta}^{\dot{\alpha}}\partial_m
+\frac{1}{4}\theta^2\bar{\theta}^2\Box)A(x)+\sqrt{2}\theta\psi(x)-
\frac{i}{\sqrt{2}}\theta^2\partial_{\alpha\dot{\alpha}}\psi^{\alpha}(x)
\bar{\theta}^{\dot{\alpha}}+
\theta^2 F(x)
\end{equation}
   The antichiral field $\bar{\sigma}$ can be obtained by conjugation.

Supersymmetric covariant derivatives have the standard form
\begin{eqnarray}
D_{\alpha}=\frac{\partial}{\partial\theta^{\alpha}}+
i\sigma_{\alpha\dot{\alpha}}^m \partial_m\bar{\theta}^{\dot{\alpha}}\\
\bar{D}_{\dot{\alpha}}=-(\frac{\partial}{\partial\bar{\theta}^{\dot{\alpha}}}+
i\theta^{\alpha}\sigma_{\alpha\dot{\alpha}}^m\partial_m)\nonumber
\end{eqnarray}

Taking in the account the standard properties of Berezin integral we can write
\begin{eqnarray}
S&=&\int d^4 x((-\frac{1}{2}\frac{Q^2}{{(4\pi)}^2}\sigma\Box\bar{\sigma}+
\frac{m^2}{2}e^{\sigma+\bar{\sigma}}+\xi_1\bar{D}^{\dot{\alpha}}\bar{\sigma}
D^{\alpha}\sigma\partial_{\alpha\dot{\alpha}}(\sigma+\bar{\sigma})+\\
&+&\xi_2\bar{D}^{\dot{\alpha}}\bar{\sigma} D^{\alpha}\sigma
\bar{D}_{\dot{\alpha}}\bar{\sigma}D_{\alpha}\sigma)
\big|_{\theta^2\bar{\theta}^2}+
(\Lambda e^{3\sigma}\big|_{\theta^2}+h.c.))\nonumber
\end{eqnarray}
Here $\big|_{\theta^2\bar{\theta}^2}$, $\big|_{\theta^2}$,
$\big|_{\bar{\theta}^2}$ denotes the corresponding components of the
superfield.

   Introducing the Dirac spinors
\begin{equation}
\Psi=\left(
\begin{array}{c}
    \psi^{\alpha}\\
    \bar{\psi}_{\dot{\alpha}}
\end{array}
\right)
\end{equation}
we write the kinetic term $\frac{Q^2}{2{(4\pi)}^2}\sigma\Box\bar{\sigma}$
as follows
\begin{equation}
\label{kinet}
\frac{Q^2}{2{(4\pi)}^2}(A\Box^2 \bar{A}+F\Box \bar{F}+
i\bar{\Psi}\Box\gamma^{\mu}\partial_{\mu}\Psi)
\end{equation}
We see the component field $F$ playing a role of auxiliary field in standard
Wess-Zumino model becomes to be dynamical.

For the exponential terms one obtains
\begin{eqnarray}
\label{expo}
\Lambda e^{3\sigma}\big|_{\theta^2}+h.c.&=&
3\Lambda e^{3A}(F-3\bar{\Psi} P_{+}\Psi)+h.c.\nonumber\\
\frac{m^2}{2}e^{\sigma+\bar{\sigma}}\big|_{\theta^2\bar{\theta}^2}&=&
\frac{m^2}{2}e^{A+\bar{A}}(2\partial_m(A+\bar{A})\partial^m(A+\bar{A})+F\bar{F}
+\frac{1}{4}\Box(A+\bar{A})-\nonumber\\
&-&F\bar{\Psi} P_{-}\Psi-
\bar{F}\bar{\Psi} P_{+}\Psi-
i\bar{\Psi}\gamma^m\partial_m\Psi+\bar{\Psi}^2\Psi^2+\\
&+&4i\bar{\Psi}\gamma^{m}\Psi
(\partial_m A+ \partial_m\bar{A})\nonumber
\end{eqnarray}

For the $V_{1,2}$-vertices one obtains
\begin{eqnarray}
\label{ver1}
\xi_1\bar{D}^{\dot{\alpha}}\bar{\sigma}D^{\alpha}\sigma
\partial_{\alpha\dot{\alpha}}(\sigma+\bar{\sigma})
\big|_{\theta^2\bar{\theta}^2}&=&
\xi_1(4i\Box A\bar{\Psi}\gamma^m\partial_m P_{+}\Psi
-4i\Box\bar{A}\bar{\Psi}\gamma^m\partial_m P_{-}\Psi+\nonumber\\
&+&\partial_m A\partial^m A\Box \bar{A}+
\partial_m\bar{A}\partial^m\bar{A}\Box A+\nonumber\\
&+&\partial^m A\partial_m\bar{A}\Box(A+\bar{A})+\\
&+&4i\partial_m\bar{\Psi}\gamma^n\partial_n\partial^m \bar{A}P_{+}\Psi-
4i\partial_m \bar{\Psi}\gamma^n\partial_n\partial^m A P_{-}\Psi+\nonumber\\
&+&8i\partial_m A\partial^m\bar{\Psi}\gamma^n P_{+}\partial_n\Psi
+8i\partial_m\bar{A}\partial^m\bar{\Psi}\gamma^n
P_{-}\partial_n\Psi+\nonumber\\
&+&4F\partial_m A\partial^m \bar{F}+
4\bar{F}\partial_m\bar{A}\partial^m F+
4F\bar{F}\Box (A+\bar{A})+\nonumber\\
&+&8\bar{\Psi}\partial^m \bar{F}P_{+}\partial_m\Psi+
8\bar{\Psi}\partial_m F P_{-}\partial^m\Psi+\nonumber\\
&+&4\partial_m\bar{\Psi} P_{+}\partial^m\Psi\bar{F}+
4\partial_m\bar{\Psi}\partial^m P_{-}\Psi F+\nonumber\\
&+&8(\bar{F}\Box\bar{\Psi} P_{+}\Psi+
F\Box\bar{\Psi} P_{-}\Psi))
\nonumber
\end{eqnarray}

\begin{eqnarray}
\label{ver2}
\xi_2\bar{D}_{\dot{\alpha}}\bar{\sigma}\bar{D}^{\dot{\alpha}}\bar{\sigma}
D^{\alpha}\sigma D_{\alpha}\sigma\big|_{\theta^2\bar{\theta}^2}&=&
\xi_2(16{(\partial_m A)}^2{(\partial_n\bar{A})}^2
+16(\partial_m A)(\partial^m\bar{A}) F\bar{F}+\nonumber\\
&+&56{(\bar{\Psi}\gamma^m\Psi)}^2+
72i(\bar{\Psi}\gamma^m\partial_m\Psi)F\bar{F}+\\
&+&20i(\bar{\Psi}\gamma^m P_{+}\Psi_{\alpha}\partial_m
\bar{A}\Box A-\bar{\Psi}{\gamma}^m P_{-}\Psi\partial_n A\Box\bar{A})+
\nonumber\\
&+&72i(\bar{\Psi}{\gamma}^n P_{+}\Psi\partial_n\bar{F}F-
\bar{\Psi}\gamma^n P_{-}\Psi\partial_n F\bar{F}))\nonumber
\end{eqnarray}

As a result we have the total action in the form
\begin{eqnarray}
S&=&\int d^4x\big\{\frac{1}{2}\frac{Q^2}{16\pi^2}(A\Box^2 \bar{A}+
F\Box \bar{F}+
i\bar{\Psi}\gamma^m\partial_m\Box\Psi)+\nonumber\\
&+&3\Lambda e^{3A}(F-\bar{\Psi} P_{+}\Psi)+3\bar{\Lambda}
e^{3\bar{A}}(\bar{F}-\bar{\Psi} P_{-}\Psi)+\nonumber\\
&+&\frac{m^2}{2}e^{A+\bar{A}}(2\partial_m(A+\bar{A})\partial^m(A+\bar{A})+
F\bar{F}
+\frac{1}{4}\Box(A+\bar{A})-\nonumber\\
&-&F\bar{\Psi} P_{-}\Psi-
\bar{F}\bar{\Psi} P_{+}\Psi-
i\bar{\Psi}\gamma^m\partial_m\Psi+\bar{\Psi}^2\Psi^2+\nonumber\\
&+&4i\bar{\Psi}\gamma^{m}\Psi
(\partial_m A+ \partial_m\bar{A})\nonumber\\
&+&g_1(\frac{1}{2}i\Box A\bar{\Psi}\gamma^m\partial_m P_{+}\Psi
-\frac{1}{2}i\Box\bar{A}\bar{\Psi}\gamma^m\partial_m P_{-}\Psi+\nonumber\\
&+&\frac{1}{8}(\partial_m A\partial^m A\Box \bar{A}+\partial_m\bar{A}\partial^m
\bar{A}\Box A+
\partial_m A\partial_m\bar{A}\Box(A+\bar{A}))+\\
&+&\frac{1}{2}i\partial_m\bar{\Psi}\gamma^n\partial_n\partial^m \bar{A}P_{+}
\Psi-
\frac{1}{2}i\partial_m \bar{\Psi}\gamma^n\partial_n\partial^m A P_{-}\Psi+
\nonumber\\
&+&i\partial_m A\partial^m\bar{\Psi}\gamma^n P_{+}\partial_n\Psi
+i\partial_m\bar{A}\partial^m\bar{\Psi}\gamma^n P_{-}\partial_n\Psi+\nonumber\\
&+&\frac{1}{2}F\partial_m A\partial^m \bar{F}+
\frac{1}{2}\bar{F}\partial_m\bar{A}\partial^m F+
\frac{1}{2}F\bar{F}\Box (A+\bar{A})+
\Psi \partial^m \bar{F}P_{+}\partial_m\Psi+\nonumber\\
&+&\bar{\Psi}\partial_m F P_{-}\partial^m\Psi+
\frac{1}{2}\partial_m\bar{\Psi} P_{+}\partial^m\Psi\bar{F}+
\frac{1}{2}\partial_m\bar{\Psi}\partial^m P_{-}\Psi F+\nonumber\\
&+&(\bar{F}\Box\bar{\Psi} P_{+}\Psi+
F\Box\bar{\Psi} P_{-}\Psi))
\nonumber\\
&+&g_2[{(\partial_m A)}^2{(\partial_n\bar{A})}^2
+(\partial_m A)(\partial^m\bar{A}) F\bar{F}\nonumber\\
&+&\frac{7}{2}{(\bar{\Psi}\gamma^m\Psi)}^2+
\frac{9}{2}i(\bar{\Psi}\gamma^m\partial_m\Psi)F\bar{F}
+\frac{5}{4}i(\bar{\Psi}\gamma^m P_{+}\Psi\partial_m \bar{A}\Box A
-\bar{\Psi}\gamma^m P_{-}\Psi\partial_n A\Box\bar{A})\nonumber\\
&+&\frac{9}{2}i(\bar{\Psi}\gamma^n P_{+}\Psi\partial_n\bar{F}F-
\bar{\Psi}\gamma^n P_{-}\Psi\partial_n F \bar{F})]\big\}\nonumber
\end{eqnarray}
where we have denoted
$$g_1=8\xi_1$$
$$g_2=16\xi_2$$
Here
$$P_{+}=\frac{1}{2}(1+\gamma_5)$$
$$P_{-}=\frac{1}{2}(1-\gamma_5)$$
are the chiral projectors.

\end{document}